\documentclass[aps,prl,twocolumn,longbibliography,superscriptaddress,reprint]{revtex4-2}
\usepackage{graphicx}
\usepackage{dcolumn}
\usepackage{bm}
\usepackage{verbatim}
\usepackage{subfigure}
\usepackage{multirow}
\usepackage{setspace}
\usepackage{physics}
\usepackage{xcolor}
\usepackage{siunitx}
\usepackage{CJK}
\usepackage{mathptmx}
\usepackage{booktabs}
\usepackage[colorlinks,linkcolor=blue,anchorcolor=blue, citecolor=blue,urlcolor=blue,]{hyperref}

\begin{document}
\begin{CJK*}{UTF8}{bsmi}
\title{A Unified Spin-Fermion Framework for Magnetic Diversity in Chromium Monopnictides}
\author{Ruoshi Jiang}
\altaffiliation{corresponding email: ruoshijiang@gmail.com}
\affiliation{Department of Materials Science and Metallurgy, University of Cambridge, Cambridge CB3 0FS, United Kingdom}
\author{Bartomeu Monserrat}
\altaffiliation{corresponding email: bm418@cam.ac.uk}
\affiliation{Department of Materials Science and Metallurgy, University of Cambridge, Cambridge CB3 0FS, United Kingdom}

\begin{abstract}
Isoelectronic compounds are generally expected to exhibit related electronic, structural, and magnetic behavior. 
Chromium monopnictides CrX (X= Sb, As, P), however, display a striking evolution from high-temperature altermagnetism in NiAs-type CrSb, through double-helical antiferromagnetism in MnP-type CrAs, to the absence of resolved long-range magnetic order in MnP-type CrP. 
Here, combining first-principles calculations in the non-collinear Curie-paramagnetic state, structural analysis, magnetic phase-diagram calculations, and exchange-parameter extraction, we establish a unified microscopic framework for this evolution. 
All three compounds share a formal high-spin \(d^5\) configuration, nominally Cr\(^{1+}\), coupled to itinerant states of predominantly pnictogen character. 
From Sb to P, chemical pressure enhances ligand itinerancy and drives the structural evolution from the high-symmetry NiAs lattice to increasingly distorted MnP-type networks. 
The accompanying reconstruction of competing Cr--Cr exchanges favors A-type antiferromagnetism in CrSb and double-helical order in CrAs, while placing CrP near a frustrated regime with closely competing magnetic tendencies.
Our results establish chemical-pressure-driven exchange reconstruction as a general mechanism through which a common spin--fermion electronic structure can generate contrasting magnetic phases in strongly correlated materials.
\end{abstract}
\flushbottom
\maketitle
\end{CJK*}

Isovalent substitution provides a widely used route to chemical-pressure tuning by varying ionic size while preserving the nominal electron count, and is generally expected to produce related electronic, structural, and magnetic behavior across a material series\,\cite{Lai2025AV3Sb5,Mochizuki2004RTiO3,Zhang2023R3Ni2O7}. 
In many strongly correlated materials, however, such nominally charge-neutral tuning can drive qualitatively distinct electronic and magnetic phases\,\cite{Hu2016BaFe2AsP,Nakatsuji2000CaSrRuO4,Steppke2013YbNi4PAs}. 
This sensitivity becomes particularly pronounced when substitution occurs directly within the ligand network\,\cite{Subedi2008FeChalcogenides}. 
Understanding how the resulting intertwined structural and electronic effects select among competing low-energy phases is a central unresolved problem.

Chromium monopnictides CrX (X= Sb, As, P) provide a striking realization of this problem. 
CrSb crystallizes in the high-symmetry NiAs-type structure and develops A-type collinear antiferromagnetic order below $T_{\mathrm N}\sim700$~K\,\cite{Takei1963CrSb}. 
This state has recently attracted great interest as a $g$-wave altermagnet exhibiting large spin splitting near the Fermi level\,\cite{CrSb_altermagnet, Long2026CrSb}. 
CrAs instead adopts the lower-symmetry MnP-type structure and undergoes a first-order transition into a double-helical antiferromagnetic state below $T_{\mathrm N}\sim265$~K\,\cite{Watanabe1969Magnetic,Kallel1974Helimagnetism,Qin_CrAs_INS}. 
Moderate external pressure suppresses its coupled structural and magnetic order and induces superconductivity\,\cite{Wu2014Superconductivity, Matsuda2018Evolution}. 
CrP is isostructural with CrAs but remains a paramagnetic metal without detectable long-range magnetic order down to low temperature\,\cite{CrP_magnetism,Niu2019Nonsaturating,Selte1972CrP,Selte1975CrPAs}. 
The isoelectronic CrX series thus spans high-temperature altermagnetism, helimagnetism proximate to pressure-induced superconductivity, and paramagnetism without long-range order, providing a rich platform to explore how chemical pressure reorganizes competing electronic and magnetic degrees of freedom.
Previous electronic structure calculations have recognized the strong evolution of magnetism across the CrX series\,\cite{Ito2007CrX}, while neutron-scattering studies have described the double-helical order of CrAs in terms of competing exchange interactions\,\cite{Matsuda2018Evolution,Kallel1974Helimagnetism,Qin_CrAs_INS}. 
Nevertheless, a unified microscopic framework linking the local electronic structure across the series, the NiAs-to-MnP structural distortion, and the resulting magnetic diversity is still lacking. 
Resolving this question is essential for understanding how chemical pressure reorganizes competing exchange interactions to produce commensurate order, helical order, and paramagnetism, while providing a microscopic basis for investigating the altermagnetism and pressure-induced superconductivity in this family.

In this Letter, we develop a unified microscopic picture of the structural and magnetic evolution across CrX based on a common spin--fermion electronic structure. 
First-principles calculations using a static noncollinear representation of the Curie-paramagnetic state reveal robust Cr local moments dominated by the $S_{\rm Cr}=5/2$ configuration, while the itinerant carriers reside primarily on the pnictogen network.
From Sb to P, chemical pressure mainly enhances Cr--pnictogen hybridization and ligand itinerancy.
Among the considered formal ionic assignments, our structural analysis identifies Cr$^{1+}$X$^{1-}$ as the configuration qualitatively consistent with both the NiAs-to-MnP structural transformation and the increasing local distortion from CrAs to CrP.
We further construct a general four-sublattice Heisenberg phase diagram, organized by dimensionless exchange ratios and applicable to materials sharing the same exchange-network topology.
The extracted exchange hierarchy places the CrX compounds along a chemical-pressure trajectory from the commensurate A-type magnetic order of CrSb, through the double helix of CrAs, to a nearly frustrated regime with competing ordering vectors in CrP.
Our work thus offers a spin-fermion framework that unifies the contrasting magnetic behavior of chromium monopnictides and provides a broader perspective on strongly correlated materials.

\textit{Electronic analysis-}
The open-shell electronic structure of the ions in transition metal oxides\,\cite{Anderson1972, Dagotto2005Complexity} often experiences intra-atomic correlation on the $10$\,eV scale and inter-atomic correlation on the $100$\,meV scale, both much stronger than room temperature.
As a result, the rich \textit{emergent} lower-energy physical behaviors can only be correctly captured when constraints on this higher-energy physics are incorporated. 
Specifically, it is essential to identify the dominant valence of the transition metal elements, and the leading fluctuation in the charge, spin, or orbital channels, to establish a solid starting point for the formulation of effective models to describe the lower-energy physics.

\begin{figure}
\centering
\includegraphics[width=\columnwidth]{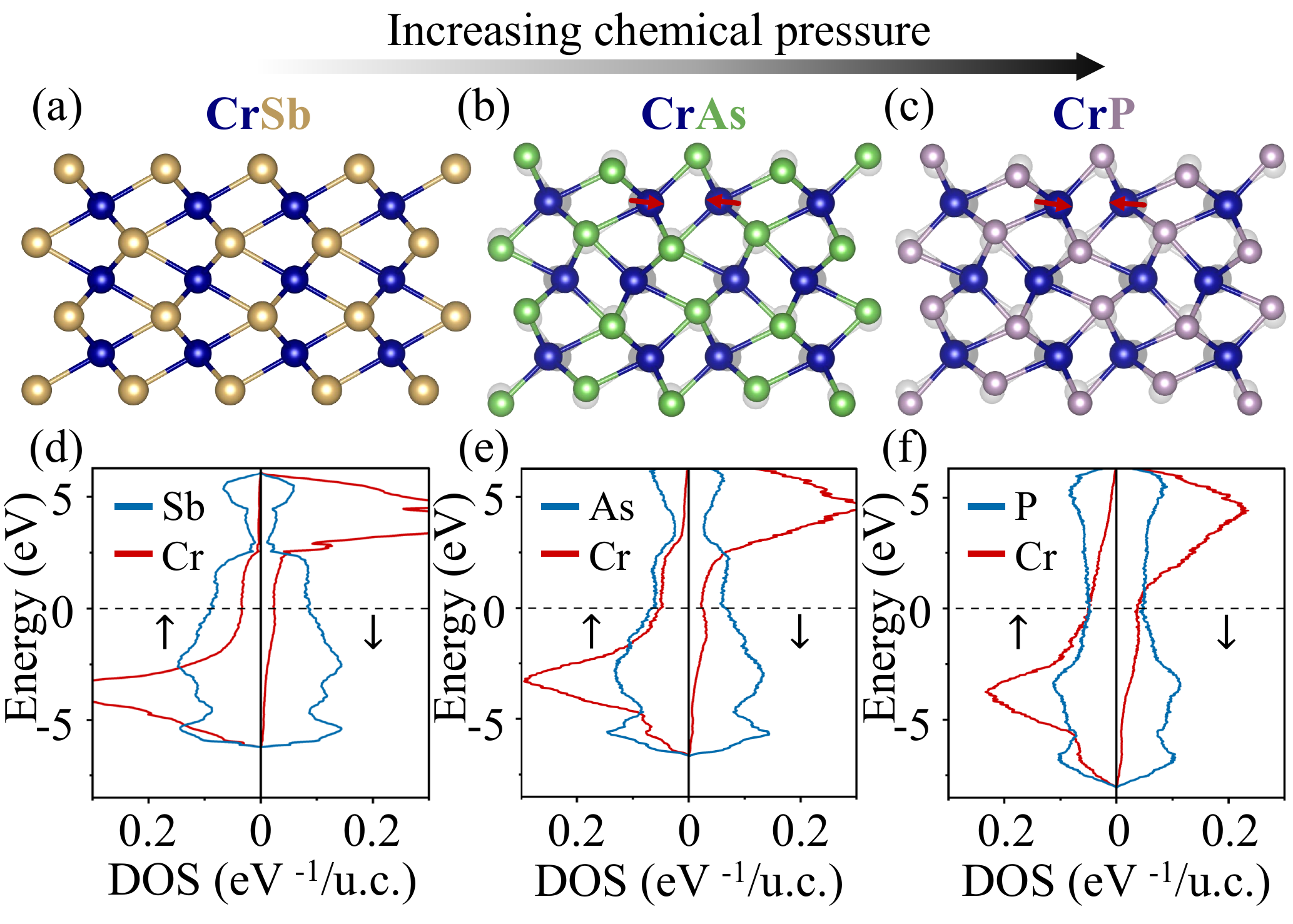}
\vspace{-0.7cm}
\caption{Chemical-pressure-driven structural and electronic evolution in CrX (X= Sb, As, P).
(a--c) NiAs-type CrSb and MnP-type CrAs and CrP viewed along the same crystallographic direction. Grey frameworks in (b)(c) indicate the idealized NiAs parent lattice; red arrows mark Cr displacements away from the high-symmetry positions.
(d--f) Corresponding spin-resolved local densities of states in the non-collinear Curie-paramagnetic state. Blue and red curves show pnictogen and Cr contributions, respectively. The Fermi level is set to zero. 
}
\label{fig:Structure}
\vspace{-0.5cm}
\end{figure}

We start by considering the Rydberg scale physics of charge distribution to identify the location of the itinerant carriers and the associated valence.
To explore this high energy regime where inter-atomic magnetic correlations are negligible, we evaluate the local density of states (LDOS) along the \textit{internal} spin orientation through simulation of the one-body spectral function in the Curie-paramagnetic phase\,\cite{Jiang2022Variation, Jiang2024Pressure} that hosts disordered \textit{non-collinear} spin orientations of the transition metal ions.
This construction removes long-range coherent magnetic order while retaining the short-range local physics, allowing large ionic moments and their associated spin-resolved spectral features to emerge if favored by the local electronic structure.

To calculate this LDOS, we use the all-electron implementation of density functional theory\,\cite{DFT1, DFT2} in {\sc Wien2k}\,\cite{Blaha2001, Blaha2019} within the LDA+$U$ formalism\,\cite{Anisimov1993,Liechtenstein1995Density-functional} with a representable $U=6$\,eV for chromium $d$-orbitals\,\cite{Tesch2022Hubbard, Wan2025High-Throughput}. 
For computational efficiency, we subsequently employ an effective many-body interacting Hamiltonian $H^{(\mathrm{Rydberg})}$\,\cite{supplementary, Jiang2022Variation, Jiang2024Pressure, Jiang2025Pressure} obtained from symmetry-preserving Wannier functions\,\cite{Wei2002Insulating, Marzari1997Maximally, supplementary}.
Within the same mean-field treatment, this many-body realistic Hamiltonian reproduces the one-body spectral function for the corresponding magnetic configurations\,\cite{supplementary} and can therefore be used to generate the Curie-paramagnetic non-collinear spin configurations of interest here.
We further verify that the dominant electronic configuration is insensitive to the choice of $U$ by applying the interaction-annealing strategy\,\cite{Jiang2024Interation}.

The resulting LDOS in the non-collinear Curie-paramagnetic states shown in Figs.~\ref{fig:Structure}(d)-(f) allows us to construct a universal picture of the dominant charge/valence profile of the chromium monopnictides CrX series.
For the transition metal Cr, shown in red, the majority-spin Cr $d$ manifold is essentially filled, whereas the minority-spin Cr $d$ states are largely unoccupied, indicating that the large intra-atomic Coulomb repulsion strongly suppresses charge fluctuations involving the transition metal ions and stabilizes a rather well-defined high-spin $d^5$ Cr configuration, nominally Cr$^{+}$ with a large local moment $5/2$.
This implies that the ligand-carrier density is broadly similar across the CrX series. 
Indeed, in great contrast to the Cr atomic orbitals, atomic orbitals of pnictogen (in blue) do not exhibit a nearly full spin polarization, indicating that the remaining carriers reside predominantly on the ligand network rather than forming rigid spin-polarized ionic moments.
This establishes a universal spin-fermion electronic structure across the series.

Consider a typical spin-fermion (SF) model consisting of coupled local moments affected by itinerant carriers\,\cite{Tam2015Itinerancy-Enhanced,Hou2024Chemical,Dagotto2012Anisotropy,Philip2010Orbital,Weng2009Coexistence},
\begin{equation}
\begin{split}
\mathcal{H}^{\rm SF} &= \sum_{ii^\prime m m^\prime \nu}t_{imi^\prime m^\prime}c_{im\nu}^\dagger c_{i^\prime m^\prime \nu} + \sum_{j\neq j^\prime}J_{jj^\prime}\mathbf{S}_j\cdot\mathbf{S}_{j^\prime}\\
&-\sum_{jim\nu i'm'\nu'}K_{jim\nu i'm'\nu'}\mathbf{S}_j c^\dagger_{im\nu}\bm{\sigma}_{\nu,\nu^\prime}c_{i' m' \nu^\prime},
\end{split}
\label{H_SpinFermion}
\end{equation}
in which itinerant carriers $c^\dagger_{im\nu}$ of spin $\nu$ in orbital $m$ of ligands located at site $i$ propagate through kinetic strength $t$.
Note that in this model, the itinerant carriers are not allowed to propagate to the transition metal ions located at site $j$, but instead their dynamics is strongly coupled to these ionic spins $\mathbf{S}_j$ via $K_{jim\nu i'm'\nu'}$, and $\mathbf{\sigma}_{\nu\nu'}$ is the standard vector of Pauli matrices.
The interplay and possible competition of carrier hopping and carrier--moment coupling with the antiferromagnetic exchange $J_{jj^\prime}$ between the ionic spins give rise to the diverse magnetic ground states across the CrX series.

Although the same formal local electronic configuration persists across the CrX series, the Cr--ligand covalency and the bandwidth of the ligand-hole states increase systematically from Sb to P [Figs.~\ref{fig:Structure}(d)--(f)]. 
This directly establishes an increasing hybridization and kinetic-energy scale under chemical pressure\,\cite{supplementary}.
These calculated trends motivate stronger itinerant renormalization of the magnetic degrees of freedom from CrSb, through CrAs, toward CrP, suggesting the possible consequences include enhanced screening and damping of the Cr moments.
The Rydberg-scale analysis therefore establishes the common local-moment and itinerant-carrier structure, but does not by itself determine whether the moments develop long-range magnetic coherence.

\textit{Structure analysis-}
To further examine whether the electronic configuration is compatible with the observed structural motifs, we analyze the corresponding coordination environments.
Figs.~\ref{fig:Structure}(a)--(c) shows the crystal structures of CrSb, CrAs, and CrP viewed along a common crystallographic direction.
CrSb adopts the high-symmetry NiAs-type structure, whereas CrAs and CrP crystallize in the orthorhombic MnP-type structure, a distorted derivative of the NiAs parent lattice indicated by the pale-gray frameworks in Figs.~\ref{fig:Structure}(b) and \ref{fig:Structure}(c).
Throughout the series, each Cr ion remains sixfold coordinated by pnictogen atoms, forming a local CrX$_6$ octahedron\,\cite{supplementary}.
The MnP distortion therefore preserves the local coordination topology while lowering its symmetry and deforming the octahedral environment.
The red arrows mark the Cr displacements from the ideal NiAs positions, illustrating the increasing local distortion from CrAs to CrP within the $Pnma$ structure\,\cite{supplementary}.

\begin{table}
\setlength{\abovecaptionskip}{3pt}
\centering
\caption{Octahedral factors $\mu=r_{\rm Cr}/r_{\rm X}$ calculated from estimated ionic radii\,\cite{shannon1976revised} for three nominal ionic configurations\,\cite{supplementary}.
The empirical stability range for regular octahedral coordination is $0.41\lesssim\mu\lesssim0.73$\,\cite{Turnley2024Rethinking}.}
\setlength{\tabcolsep}{8pt}
\vspace{5pt} 
\begin{tabular}{c|ccc}
\toprule
\hline
\rule{0pt}{2.5ex} 
$\mu = r_{\rm Cr}/r_{\rm X} $ & Cr$^{1+}$X$^{1-}$ &Cr$^{2+}$X$^{2-}$&Cr$^{3+}$X$^{3-}$ \\ \hline
\rule{0pt}{2.6ex}
CrSb & $\bm{0.67}$ & 0.41 & 0.28 \\
CrAs & $\bm{0.83}$ & 0.49 & 0.33 \\
CrP & $\bm{0.90}$ & 0.53 & 0.35 \\
\botrule
\end{tabular}
\vspace{-0.6cm}
\label{tab1_OctahedralFactor}
\end{table}

To provide a qualitative ionic-size perspective on this structural evolution, we evaluate the octahedral factor $\mu=r_{\rm Cr}/r_{\rm X}$, for which the classical radius-ratio criterion favors sixfold coordination within the approximate window $0.414\lesssim\mu\lesssim0.732$\,\cite{Chonghe2004Formability,Turnley2024Rethinking}. 
Using effective ionic radii\,\cite{supplementary,shannon1976revised}, we compare three nominal valence scenarios in Table~\ref{tab1_OctahedralFactor}. 
The Cr$^{3+}$X$^{3-}$ scenario yields $\mu$ values well below this window for all three compounds, whereas Cr$^{2+}$X$^{2-}$ places them all within it, providing no clear distinction between the undistorted NiAs-type CrSb and the distorted MnP-type CrAs and CrP. 
In contrast, the Cr$^{+}$X$^{-}$ scenario places CrSb within the window, while CrAs and especially CrP exceed its upper boundary, suggesting an increasing ionic-size mismatch as the ligand size decreases. 
Within this approximate ionic picture, this trend qualitatively parallels the observed symmetry lowering and increasing distortion of the CrX$_6$ environment in the CrX series, providing a structural consistency check for the Cr $d^5$ configuration inferred from the Rydberg-scale LDOS analysis above.

\textit{Magnetic analysis-}
To understand how chemical pressure reorganizes the magnetic properties across the CrX series, we further integrate out the degrees of freedom of itinerant carriers and retain only an effective description in terms of magnetic correlations among the renormalized Cr local moments. 
Since the MnP-type distortion lowers the symmetry of the Cr sublattice and splits the Cr--Cr exchange network into several inequivalent pathways, we build a four-sublattice Heisenberg model\,\cite{Kallel1974Helimagnetism,Matsuda2018Evolution,Qin_CrAs_INS} that retains the dominant symmetry-inequivalent Cr--Cr exchange interactions,
\begin{equation}
\mathcal{H}
=\sum_{\langle ij\rangle}
\tilde{J}_{ij}
\tilde{\mathbf{S}}_i \cdot \tilde{\mathbf{S}}_j ,
\end{equation}
where $\tilde{\mathbf{S}}_i$ denotes the renormalized Cr spin on sublattice $i$. 
Positive and negative $\tilde{J}_{ij}$ correspond to antiferromagnetic (AFM) and ferromagnetic (FM) interactions, respectively.

\begin{figure}
\centering
\includegraphics[width=1\columnwidth]{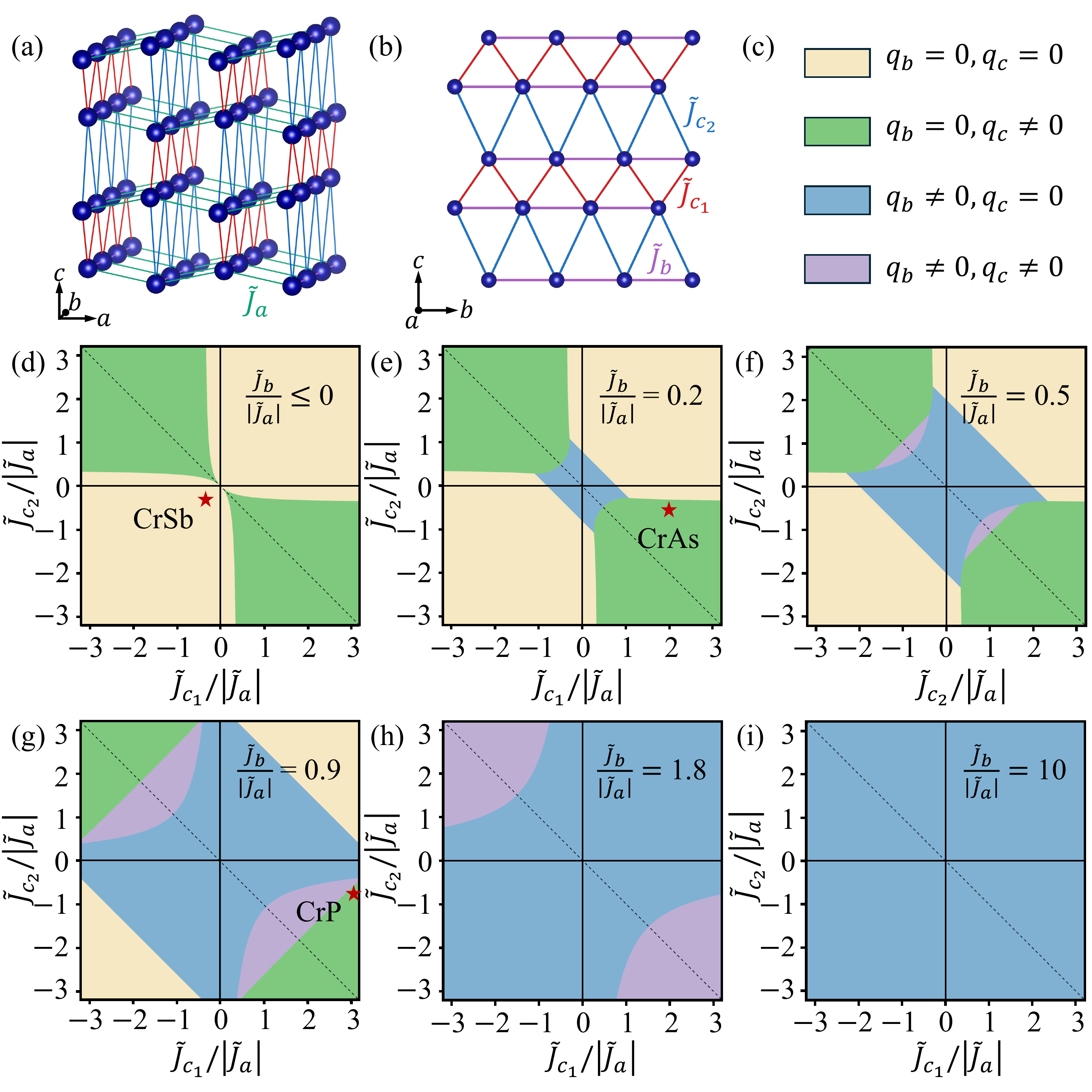}
\vspace{-0.7cm}
\caption{Exchange network and classical ground-state phase diagrams of the effective four-sublattice Heisenberg model.
(a) Three-dimensional Cr exchange network, with $\tilde{J}_a$, $\tilde{J}_b$, $\tilde{J}_{c_1}$, and $\tilde{J}_{c_2}$ shown in green, purple, red, and blue, respectively.
(b) Projection onto the $bc$ plane, highlighting triangular motifs formed by $\tilde{J}_b$, $\tilde{J}_{c_1}$, and $\tilde{J}_{c_2}$.
(c) Color key classifying the ground states according to whether the components $q_b$ and $q_c$ are zero or finite.
(d)--(i) Phase diagrams in the $\tilde{J}_{c_1}/|\tilde{J}_a|,\tilde{J}_{c_2}/|\tilde{J}_a|$ plane for the indicated $\tilde{J}_b/|\tilde{J}_a|$.
Red stars indicate the estimated exchange ratios for CrSb, CrAs, and CrP. 
}
\label{fig:phase-slices-main}
\vspace{-0.6cm}
\end{figure}

Figure~\ref{fig:phase-slices-main}(a) depicts the three-dimensional Cr exchange network, with the green, purple, red, and blue bonds denoting $\tilde{J}_a$, $\tilde{J}_b$, $\tilde{J}_{c_1}$, and $\tilde{J}_{c_2}$, respectively.
Figure~\ref{fig:phase-slices-main}(b) shows its projection onto the $bc$ plane, where $\tilde{J}_b$, $\tilde{J}_{c_1}$, and $\tilde{J}_{c_2}$ form triangular exchange motifs that can impose mutually competing spin constraints, depending on their relative signs and magnitudes.
Within this four-sublattice representation, the higher-symmetry NiAs structure is recovered in the symmetry-constrained limit
$\tilde{J}_{c_1}=\tilde{J}_{c_2}=\tilde{J}_b$.

We examine how exchange competition selects the magnetic ordering vector by treating the large effective Cr moments classically and minimizing the lowest eigenvalue of the exchange matrix. 
This minimization fixes $q_a=0$ since the retained $a$-direction exchange does not support a competing finite-$q_a$ modulation\,\cite{supplementary}. 
The ground-state ordering vector therefore takes the form $\mathbf q=(0,q_b,q_c)$, with $q_b=\mathbf q\cdot\mathbf b$ and $q_c=\mathbf q\cdot\mathbf c$ denoting the intercell phase advances along $b$ and $c$, respectively. 
For a single-$\mathbf q$ state, $q_\alpha=0$ and $q_\alpha=\pi$ ($\alpha=b,c$) denote repetition and sign reversal of the spin pattern between neighboring unit cells, respectively, whereas intermediate values $0<q_\alpha<\pi$ describe a spiral phase advance along the corresponding direction. 

The calculated phase diagrams in Figs.~\ref{fig:phase-slices-main}(d)--(i) show the evolution of the optimal ordering vector in the $(\tilde{J}_{c_1}/|\tilde{J}_a|,\tilde{J}_{c_2}/|\tilde{J}_a|)$ plane for several representative values of $\tilde{J}_b/|\tilde{J}_a|$ under the color key classifying the ground states according to whether the components $q_b$ and $q_c$ are zero or finite shown in Fig.~\ref{fig:phase-slices-main}(c). 
The corresponding maps of $q_b$ and $q_c$ are provided in the Supplemental Material\,\cite{supplementary}. 
At $\tilde{J}_b=0$, the first and third quadrants satisfy $\tilde{J}_{c_1}\tilde{J}_{c_2}>0$, such that the two $c$-related exchange paths are either both AFM or both FM and favor $q_c=0$.
In the second and fourth quadrants where $\tilde{J}_{c_1}\tilde{J}_{c_2}<0$, by contrast, their opposite signs impose competing inter-sublattice exchange constraints that can stabilize a finite $q_c$.
The stability and extent of these finite-$q_c$ regions are governed by their relative exchange strengths\,\cite{supplementary}.

The role of $\tilde{J}_b$ is qualitatively different.
For $\tilde J_b<0$, FM exchange along $b$ favors $q_b=0$.
In this sector, $\tilde J_b$ contributes a $q_c$-independent energy shift, so varying its magnitude does not change the phase boundaries, consistent with the common $\tilde J_b/|\tilde J_a|\leq 0$ cut in Fig.~\ref{fig:phase-slices-main}(d).
For $\tilde{J}_b>0$, the exchange favors AFM correlations along $b$ and competes with $\tilde{J}_{c_1}$ and $\tilde{J}_{c_2}$ within the triangular exchange motifs.
Increasing $\tilde{J}_b/|\tilde{J}_a|$ therefore drives the system toward a staggered modulation along $b$ while progressively suppressing the finite-$q_c$ regions.
At larger $\tilde{J}_b/|\tilde{J}_a|$, $q_b$ approaches $\pi$, and a finite $q_c$ is no longer stabilized within the parameter range considered.

We emphasize that $q_b=q_c=0$ specifies only the translational periodicity and does not necessarily imply a FM ground state.
The ordering vector determines the phase advance between crystallographic unit cells, whereas the relative orientations of the four spins within each unit cell are encoded in the eigenvector of the lowest exchange band.
Consequently, a $\mathbf q=0$ state repeats the same four-sublattice magnetic motif in every unit cell, but this intracell motif may be ferromagnetic, antiferromagnetic, or even a noncollinear arrangement.

Importantly, these phase diagrams are not specific to the CrX series.
Because the ground states are organized by the dimensionless ratios $\tilde{J}_b/|\tilde{J}_a|$, $\tilde{J}_{c_1}/|\tilde{J}_a|$, and $\tilde{J}_{c_2}/|\tilde{J}_a|$, they provide a general classification for magnets sharing the same four-sublattice exchange connectivity, irrespective of their absolute exchange scale or moment size at the classical level.
This framework therefore extends to a broader family of MnP-type transition-metal compounds, including MnP, FeP, and FeAs, which collectively realize commensurate, helical, and other noncollinear magnetic states\,\cite{Forsyth1966The,Kallel1974Helimagnetism,Chernyavskii2020Incommensurate,Sukhanov2022Frustration,Rodriguez2011Noncollinear}, as well as paramagnetic members such as CoP, and CoAs\,\cite{Kallel1974Helimagnetism, Stein1966Monophosphides, Campbell2018CoAs}.
In MnP, pressure further changes the preferred helical propagation direction from the $c$ axis to the $b$ axis, providing a particularly direct example of exchange-driven axis selection\,\cite{Matsuda2016Pressure,Khasanov2016High-pressure,Matsuda2018Polarized,Dissanayake2023Helical}.
Pressure, strain, or chemical substitution can thus be viewed as driving material-specific trajectories through this common exchange-parameter space, allowing distinct magnetic phases to emerge from the same underlying frustrated network\,\cite{Cheng2015Pressure-Induced,Xu2017First-Principles,Wang2016Spiral,Kallel1974Helimagnetism}.
Longer-range exchange, magnetic anisotropy, and spin--orbit interactions can shift the phase boundaries and may modify the detailed ordering selection\,\cite{Matsuda2018Polarized,Sukhanov2022Frustration}. 
Within the minimal four-exchange model, however, the inequivalent \(b\)- and \(c\)-direction pathways are already sufficient to generate several competing ordering minima and to demonstrate how structural reconstruction changes the preferred propagation vector.

\begin{table}[t]
\setlength{\abovecaptionskip}{3pt}
\centering
\caption{Dominant effective local moments and Cr--Cr exchange parameters (meV) up to fourth-nearest neighbors in CrX, obtained from TB2J using an A-type AFM reference\,\cite{supplementary}.
Positive (negative) values denote AFM (FM) exchange.}
\setlength{\tabcolsep}{8pt}
\vspace{5pt}
\begin{tabular}{c|c|cccc}
\toprule \hline
\rule{0pt}{2.5ex}
& $\tilde{S}$ & $|\tilde{J}_{a}|$ & $\tilde{J}_{b}$ & $\tilde{J}_{c_1}$ & $\tilde{J}_{c_2}$ \\ \hline 
\rule{0pt}{2.5ex}
CrSb & 2.06 & 7.32 & $-$2.19 & $-$2.19 & $-$2.19\\ 
\rule{0pt}{2.5ex}
CrAs & 1.84 & 9.38 & 1.91 & 18.98 & $-$4.60 \\
\rule{0pt}{2.5ex}
CrP  & 1.67 & 10.34 & 9.06 & 31.86 & $-$8.10 \\ \hline
\bottomrule
\end{tabular}
\vspace{-0.5cm}
\label{tab:ExchangeCoupling}
\end{table}

To assess how itinerant carriers modify the effective interactions among Cr local moments across the CrX family, we estimate the magnetic exchange couplings using the TB2J procedure\,\cite{Liechtenstein1987Local,He2021TB2J}.
We note that the individual exchange couplings obtained from TB2J depend appreciably on the chosen magnetic reference states\,\cite{He2021TB2J,Zhu2020Magnetic,Szilva2023Quantitative,supplementary} and should therefore not be regarded as uniquely determining the magnetic ground state.
This limitation is particularly important when the estimated parameters lie close to a phase boundary, where modest variations in the couplings can change the predicted ordering vector.
Nevertheless, their overall hierarchy and systematic evolution across the CrX series remain informative, revealing how chemical pressure reshapes the competition among exchange pathways and the resulting magnetic instabilities.

Table~\ref{tab:ExchangeCoupling} presents a representative set of exchange parameters obtained from the A-type AFM reference state, retaining the dominant interactions up to the fourth-nearest-neighbor Cr pairs.
This choice is motivated by the experimentally established A-type AFM ground state of CrSb\,\cite{Takei1963CrSb}.
Applying the same configuration for CrAs and CrP avoids introducing additional compound-dependent bias associated with using different reference states, allowing the evolution of the exchange couplings to be compared consistently as a function of chemical pressure.
Results obtained from other magnetic reference states are summarized in the Supplemental Material\,\cite{supplementary}.

Within the common A-type reference, the effective Cr spin amplitude decreases from $\tilde{S}\simeq2.06$ in CrSb to $1.84$ in CrAs and $1.67$ in CrP. 
Its reduction from the ionic $S=5/2$ value in Eq.~\eqref{H_SpinFermion} is consistent with renormalization by increasing kinetic energy of itinerant carriers.
The dominant exchange scale grows concurrently, consistent with stronger superexchange from enhanced hopping ($\sim t^2/U$ in the strong-coupling limit).
The truncated exchange hierarchy places CrSb in a commensurate AFM regime.
CrAs favors a helical state with $q_c/2\pi\simeq0.37$, close to the experimentally observed propagation vector of its double helix\,\cite{Qin_CrAs_INS}.
CrP also lies in the finite-$q_c$ helical regime, but its enhanced AFM $\tilde{J}_b$ strengthens exchange competition within the $bc$ plane, placing it near the boundary with a phase in which both $q_b$ and $q_c$ are finite.

The effective Cr spin amplitudes depend only weakly on the magnetic reference state, whereas the extracted exchange parameters exhibit appreciable sensitivity\,\cite{supplementary}. 
Accordingly, only qualitative trends that persist across these reference states should be regarded as robust, instead of overinterpreting the absolute values of individual exchange parameters and the corresponding wave vector.
The comparison with experiment remains an essential benchmark.

Consistent with the experiments of CrSb\,\cite{Takei1963CrSb, Singh2026Coherent}, the couplings extracted from the A-type reference yield an AFM $\tilde J_a$ between adjacent Cr planes and FM inplane magnetic coupling.
For CrAs, this exchange hierarchy obtained independently from our calculations reproduce the key experimental hierarchy from available inelastic-neutron-scattering\,\cite{Qin_CrAs_INS}, particularly the competing AFM $\tilde{J}_{c_1}$ and FM $\tilde{J}_{c_2}$.
Our calculations, however, yield the opposite sign of $\tilde{J}_a$ from that inferred by the neutron-scattering fit\,\cite{Qin_CrAs_INS}, corresponding to an intracell $\pi$ phase shift between sublattices along $a$ direction.
This overall agreement supports the A-type configuration as a common reference state and lends credibility to the qualitative exchange trends predicted for CrP.
In the absence of experimental exchange parameters for CrP, the most robust prediction is an enhanced AFM $\tilde J_b$ and stronger competition among ordering-vector sectors\,\cite{supplementary}, which could be directly tested by inelastic neutron scattering.

Within our spin--fermion picture, the suppression of helimagnetism in CrAs by pressure and P substitution, despite persistent short-range AFM correlations\,\cite{Matsuda2018Evolution}, is consistent with the loss of long-range coherence without extinguishing local magnetic correlations. 
Moreover, the enhanced exchange competition and reduced Cr spin amplitude in CrP are consistent with enhanced magnetic fluctuations and itinerant renormalization. 
We propose that screening and damping by the itinerant ligand-dominated carriers may further inhibit long-range phase coherence.  
This offers a possible interpretation of its weakly temperature-dependent susceptibility and absence of resolved magnetic order in terms of dynamically concealed Cr moments\,\cite{Jiang_localmoment}.

In conclusion, we establish CrX (X= Sb, As, P) as a platform in which chemical pressure reconstructs competing exchange interactions within a common spin-fermion electronic structure which encapsulates a rich phase diagram spanning commensurate order, helimagnetism, and competing incommensurate magnetism, offering a unique testbed for understanding competing low-energy physics in quantum materials.
Our results distinguish local-moment formation from long-range magnetic coherence and provide a microscopic basis for investigating how the resulting magnetic fluctuations influence transport, altermagnetism, and pressure-induced superconductivity across this material family.
More generally, they reveal how itinerancy and exchange frustration govern the long-range coherence of well-defined local moments in the strongly correlated materials.

We thank Prof. F. Malte Grosche, John Leung, Prof. Wei Ku, Dr. Fangyuan Gu, and Dr. Jiahao Yang for helpful discussions. 
This work is supported by a UKRI Future Leaders Fellowship [UKRI2083] and an EPSRC grant [EP/V062654/1].

\bibliography{main.bib}

\end{document}